\documentclass[preprint,review,12pt]{elsarticle}

\usepackage{amssymb}
\usepackage{amsmath}
\usepackage{booktabs}
\usepackage{hyperref}

\biboptions{sort&compress}

\usepackage{xcolor}
\usepackage{makecell}

\journal{Solid-State Electronics}

\begin{document}

\begin{frontmatter}



\title{Modeling Incomplete Conformality during Atomic Layer Deposition in High Aspect Ratio Structures}

\author[inst1]{Luiz Felipe Aguinsky\corref{cor1}}
\ead{aguinsky@iue.tuwien.ac.at}

\author[inst1]{Fr{\^a}ncio Rodrigues}
\author[inst2]{Tobias Reiter}
\author[inst2]{Xaver Klemenschits}
\author[inst2]{Lado Filipovic}
\author[inst3]{Andreas H{\"o}ssinger}
\author[inst1]{Josef Weinbub}

\cortext[cor1]{Corresponding author}

\affiliation[inst1]{organization={Christian Doppler Laboratory for High Performance TCAD, Institute for Microelectronics, TU Wien},
            addressline={Gußhausstraße 27-29}, 
            postcode={1040},
            city={Wien},
            country={Austria}}

\affiliation[inst2]{organization={Institute for Microelectronics, TU Wien},
            addressline={Gußhausstraße 27-29}, 
            postcode={1040},
            city={Wien},
            country={Austria}}
            
\affiliation[inst3]{organization={Silvaco Europe Ltd.},
    addressline={Compass Point}, 
    city={St Ives, Cambridge},
    postcode={PE27 5JL},
    country={United Kingdom}}

\begin{abstract}
Atomic layer deposition allows for precise control over film thickness and conformality. It is a critical enabler of high aspect ratio structures, such as 3D NAND memory, since its self-limiting behavior enables higher conformality than conventional processes. However, as the aspect ratio increases, deviations from complete conformality frequently occur, requiring comprehensive modeling to aid the development of novel technologies. To that end, we present a model for surface coverage during atomic layer deposition where incomplete conformality is present. This model combines existing approaches based on Knudsen diffusion and Langmuir kinetics. Our model expands the state-of-the art by (i) incorporating gas-phase diffusivity through the Bosanquet formula as well as reaction reversibility in the modeling framework first proposed by Yanguas-Gil and Elam, and (ii) being efficiently integrated within level-set topography simulators. The model is manually calibrated to published results of the prototypical atomic layer deposition of Al$_2$O$_3$ from TMA and H$_2$O in lateral high aspect ratio structures. We investigate the temperature dependence of the H$_2$O step, thus extracting an activation energy of $0.178\,\mathrm{eV}$ which is consistent with recent experiments. In the TMA step, we observe increased accuracy from the Bosanquet formula and we reproduce multiple independent experiments with the same parameter set, highlighting that the model parameters effectively capture the reactor conditions.
\end{abstract}



\end{frontmatter}

\section{Introduction}
\label{sec:intro}
Atomic layer deposition (ALD) is a thin film deposition technique which enables greater control over film thickness and conformality than conventional chemical vapor deposition (CVD)~\cite{cremers2019conformality}. ALD has become a key technology in semiconductor processing, having found application in, e.g., the deposition of technologically relevant oxides and nitrides~\cite{knoops2015atomicbook}. Due to its increased control over conformality, ALD is a key enabler of high aspect ratio (HAR) structures such as dynamic random-access memory (DRAM) capacitors~\cite{jakschik2004physical} and three-dimensional (3D) NAND flash memory~\cite{fischer2022control}.

In contrast to conventional CVD, ALD divides the growth process into at least two sequential, self-limiting processing steps, which repeat in cycles~\cite{knoops2015atomicbook}. From the many precursor chemistries enabling ALD, the deposition of aluminum oxide (Al$_2$O$_3$) from trimethylaluminum (TMA, or Al(CH$_3$)$_3$) and water (H$_2$O) has emerged as a paradigmatic system~\cite{puurunen2005surface}. Even though this process has found application in, e.g., high-$\kappa$ capacitor films for DRAM~\cite{jakschik2004physical}, its main importance stems from the near-ideal aspects of the involved surface chemistry. Thus, a significant body of research has emerged for this process, and it became the \textit{de facto} standard against which novel approaches are tested.

In an irreversible self-limiting reaction with fixed reactor conditions, complete conformality is theoretically achievable by adapting the step pulse time $t_p$ to the involved HAR structure. Thus, the conformal film thickness could be straightforwardly controlled via the growth per cycle (GPC) parameter, determined by the involved reactants and reactor conditions, and the total number of cycles ($N_\mathrm{cycles}$). However, in real-world conditions, deviations from complete conformality in HAR structures are observed~\cite{cremers2019conformality} since (i) the true surface chemistry is not perfectly self-limiting, and (ii) reactant transport becomes severely constricted. Accordingly, as semiconductor technology advances towards ever higher aspect ratios, the challenge of understanding incomplete conformality in ALD must be addressed with a joint experimental and modeling approach.


To that end, first-order Langmuir models have been developed and applied to predict saturation times~\cite{gobbert2002predictive,gordon2003kinetic,yanguas2012self}, to model growth kinetics~\cite{ylilammi2018modeling}, to derive scaling laws~\cite{szmyt2021atomic}, and to estimate the clean surface sticking coefficient ($\beta_0$) using either Monte Carlo methods~\cite{schwille2017temperature,poodt2017effect} or simplified analytical expressions~\cite{arts2019sticking}. These approaches are very powerful, however, they do not evaluate the resulting thickness profiles in a manner which is compatible with level-set topography simulators. This is a requirement for the integration of ALD models with additional processing steps and for process-aware device simulation within a design-technology co-optimization (DTCO) framework~\cite{TUW-297930}.

In the past, we addressed this issue in the context of the ALD of titanium compounds by developing a topography simulation integrating detailed Langmuir surface models with Monte Carlo ray tracing calculations of local reactant fluxes~\cite{filipovic2019modeling}. Nevertheless, the use of Monte Carlo ray tracing as well as the calculation of the growth on a cycle-by-cycle basis leads to high computational costs. Therefore, only a few deposition cycles were simulated. For a topography simulation of realistic ALD processes involving hundreds of cycles, not only the surface coverages but also the level-set velocity field must be accurately and efficiently calculated.

Here, we present a model for ALD surface coverage in HAR structures based on one-dimensional (1D) diffusive particle transport, building upon the model proposed by Yanguas-Gil and Elam~\cite{yanguas2012self} by combining it with physical-chemical phenomena highlighted in previous works~\cite{gobbert2002predictive,ylilammi2018modeling,yim2022conformality}. Namely, the model now includes reversible reactions and gas-phase diffusion through the Bosanquet formula~\cite{pollard1948gaseous}. For the calculation of thickness profiles, the model is efficiently integrated with level-set based topography simulators~\cite{sethian1999level,klemenschits2018modeling,ViennaLS,VictoryProcess} through the bundling of multiple cycles via the introduction of an artificial time unit. Our model is then manually calibrated to reported ALD thicknesses of Al$_2$O$_3$ in both the H$_2$O- and TMA-limited regimes, allowing for a deeper analysis of the role of temperature and geometrical parameters for this prototypical process.

\section{Methods}
\label{sec:methods}

\subsection{Surface kinetics and flux modeling}
\label{ssec:knudsenlangmuir}


As with most ALD modeling approaches~\cite{cremers2019conformality}, our model assumes that the processes are limited by the reactive transport of a single reactant species. 
For clarity, our discussion focuses on the H$_2$O-limited regime during ALD of Al$_2$O$_3$. However, the same insights are valid for the TMA-limited case and to similar reactants. We employ a first-order Langmuir surface model, combined with diffusive reactant transport for the calculation of the surface coverage $\theta$, building upon the model first proposed by Yanguas-Gil and Elam~\cite{yanguas2012self} by considering reversible kinetics and the impact of gas-phase diffusivity~\cite{gobbert2002predictive,ylilammi2018modeling,yim2022conformality}. 

\begin{figure}[h!]
    \centering 
    \includegraphics[width=0.6\columnwidth]{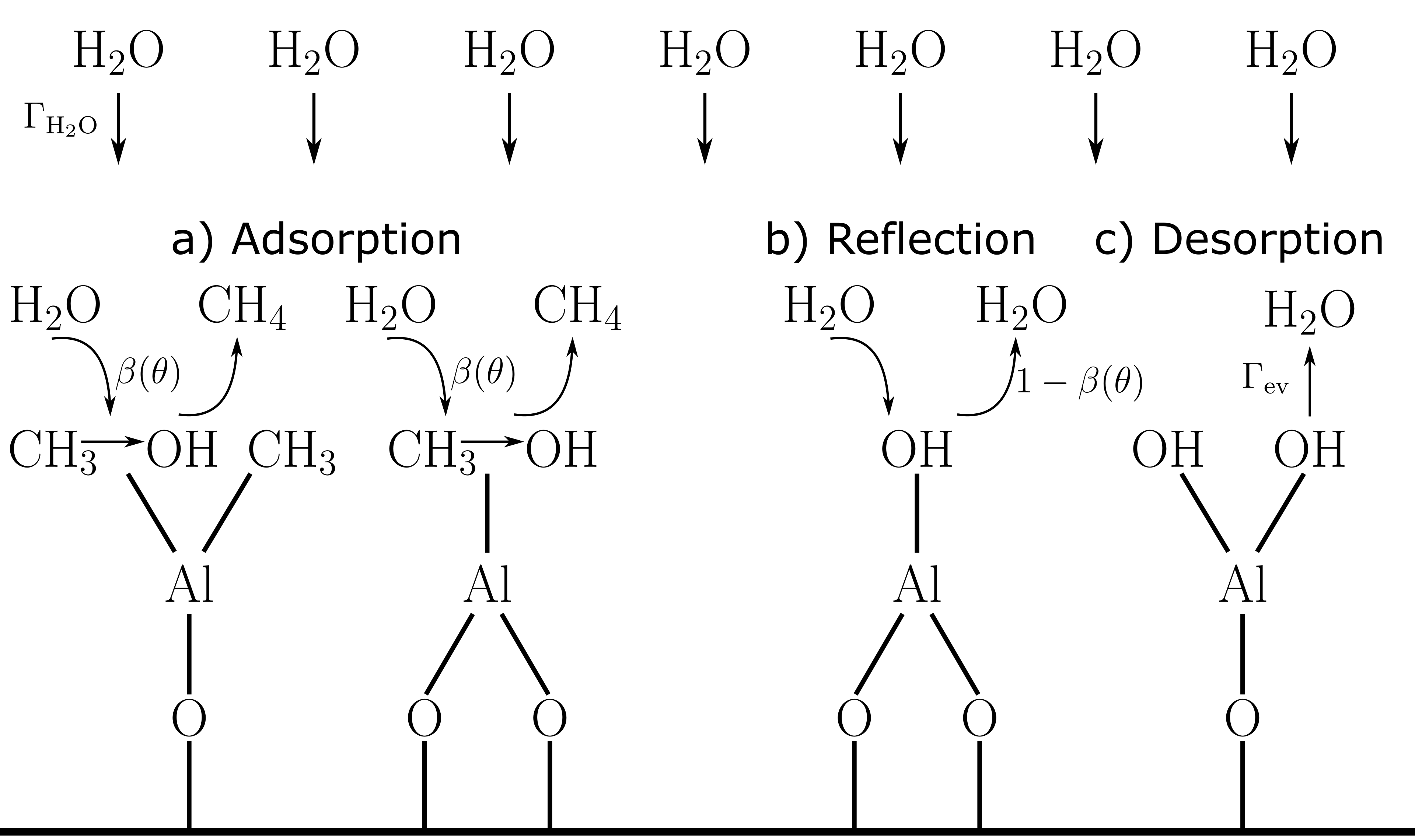}
    \caption{Possible reaction pathways in reversible Langmuir kinetics for the H$_2$O step of ALD of Al$_2$O$_3$.}  
    \label{fig:chem}
\end{figure}

The following reaction pathways for an impinging water flux $\Gamma_\mathrm{H_2O}$ ($\mathrm{m^{-2}\, s^{-1}}$) are considered, represented in Fig.~\ref{fig:chem}: Adsorption-reflection, mediated by a $\theta$-dependent sticking coefficient $\beta(\theta) = \beta_0(1-\theta)$, and desorption, given by an evaporation flux $\Gamma_\mathrm{ev}$ ($\mathrm{m^{-2}\, s^{-1}}$). In the original model~\cite{yanguas2012self} as well as in subsequent developments~\cite{keuter2015modeling,szmyt2021atomic,yanguas2021reactor}, irreversible kinetics are assumed, i.e., $\Gamma_\mathrm{ev} = 0$. However, other works have highlighted the necessity of considering the reaction reversibility, leading to the following equation for the time evolution of $\theta$ at each surface point $\vec{r}$~\cite{gobbert2002predictive,ylilammi2018modeling,yim2022conformality}:
\begin{equation}
    \label{eq:langmuir}
    \frac{1}{s_0}\frac{d\theta(\vec{r})}{dt} = \Gamma_{\mathrm{H_2O}}(\vec{r})\overbrace{\beta_0\left(1-\theta(\vec{r})\right)}^{\beta(\theta)} - \Gamma_\mathrm{ev}\theta(\vec{r})
\end{equation}

Equation~\eqref{eq:langmuir} describes an empirical model with two phenomenological parameters: $\beta_0$ and $\Gamma_\mathrm{ev}$. The surface site area $s_0$ ($\mathrm{m^2}$) can be estimated with a ``billiard ball" approximation from the deposited film density $\rho$ ($\mathrm{kg\,m^{-3}}$) and GPC ($\text{\r{A}}$)~\cite{ylilammi2018modeling}. In contrast to the steady-state assumption applied in, e.g., plasma etching simulations~\cite{klemenschits2018modeling}, we solve~\eqref{eq:langmuir} up to the reactor pulse time $t_p$ ($\mathrm{s}$) using the forward Euler method with $N_t$ total time steps. The purge step is not considered.

A requirement for determining $\theta(\vec{r})$ is finding the distribution of the reactant flux $\Gamma_{\mathrm{H_2O}}(\vec{r})$. This calculation is challenging given that the $\beta(\theta)$ changes not only across the surface but also after the solution of each step of~\eqref{eq:langmuir}. Although powerful methods such as the Boltzmann transport equation~\cite{gobbert2002predictive}, the lattice Boltzmann model~\cite{fang2019atomic}, or Monte Carlo ray tracing can be used~\cite{schwille2017temperature,filipovic2019modeling}, they require substantial computational resources.

To alleviate the computational burden, we assume a preferential transport direction, i.e., that the flux is equal on all surfaces at the same $z$ coordinate. This allows to calculate the flux using the continuity equation assuming diffusive flow in a cylinder of diameter $d$ ($\mu\mathrm{m}$) and length $L$ ($\mu\mathrm{m}$), with adsorption losses, given by a 1D differential equation~\cite{yanguas2012self,yanguas2016growth}:
\begin{align}
    D\frac{d^2\Gamma_{\mathrm{H_2O}}(z)}{dz^2} &= \frac{\bar{v}}{d}\beta_0\left(1-\theta(z)\right)\Gamma_{\mathrm{H_2O}}(z) \text{,} \nonumber\\
        \Gamma_{\mathrm{H_2O}}(0) &= \Gamma_{0} \text{,} \label{eq:diffusion}\\
        D\frac{d\Gamma_{\mathrm{H_2O}}}{dz}\Bigr|_{z=L} &= -\frac{1}{4}\bar{v}\beta_0\left(1-\theta(L)\right)\Gamma_\mathrm{H_2O}(L) 
    \nonumber
\end{align}
In \eqref{eq:diffusion}, $\bar{v}$ ($\mathrm{m\,s^{-1}}$) is the thermal speed and $\Gamma_0$ ($\mathrm{m^{-2}\,s^{-1}}$) is the flux of the reactant species inside the reactor, which can be calculated using the kinetic theory of gases~\cite{chapman1990mathematical} from the reactor temperature $T$ ($\mathrm{^\circ C}$), reactant molar mass $M_A$ ($\mathrm{kg\,mol^{-1}}$), and partial pressure $p_A$ ($\mathrm{mTorr}$). The frozen surface approximation is employed~\cite{yanguas2012self}, that is, transport is assumed to reach equilibrium on a much faster timescale than the chemical evolution of the surface (i.e., $d \Gamma_{\mathrm{H_2O}}/dt = 0$ even though $d\theta/dt \neq 0$). This approximation is generally accepted as valid for microscopic structures~\cite{yanguas2016growth}. This equation is solved with a central finite differences scheme for each step of the solution of \eqref{eq:langmuir}.

The system composed of~\eqref{eq:langmuir} and~\eqref{eq:diffusion} is, in essence, a re-statement of the established Yanguas-Gil and Elam model~\cite{yanguas2012self} with two main differences. First, the reversibility of the reactions is considered in~\eqref{eq:langmuir}. Also, the diffusivity $D$ ($\mathrm{m^2\,s}^{-1}$) is considered explicitly, which enables to combine Knudsen diffusion with gas-phase diffusion through the Bosanquet interpolation formula, which is discussed in the following paragraph. Both of these physical-chemical phenomena have been incorporated into modeling by previous studies either separately~\cite{gobbert2002predictive,poodt2017effect} or jointly~\cite{ylilammi2018modeling,yim2022conformality}. However, such models, most notably the approach taken by Ylilammi \textit{et al.}~\cite{ylilammi2018modeling} and its subsequent expansion~\cite{yim2022conformality}, rely on a different set of approximations for the calculation of the flux distribution inside the structure and do not compute a solution to~\eqref{eq:diffusion}. In their work, the frozen surface approximation is not employed and the partial pressure distribution, which is equivalent to the reactant flux distribution, is directly approximated as two separate regions, one linear and another exponentially decaying.

As previously indicated, $D$ can be calculated by considering two individual contributions: One stemming from reactant-wall collisions, i.e., the Knudsen diffusivity $D_\mathrm{Kn}$, and another stemming from reactant-reactant collisions, i.e., the gas-phase diffusivity $D_A$. Historically, Knudsen diffusion has been defined in terms of a long cylindrical tube~\cite{Knudsen1909} of diameter $d$, leading to the following expression for the diffusivity~\cite{pollard1948gaseous}:
\begin{equation}
    \label{eq:knudsen}
    D_\mathrm{Kn} = \frac{1}{3}\bar{v}d
\end{equation}

Therefore, when structures other than a long cylindrical tube are considered, some kind of mapping between the involved geometry and the standard cylinder must be provided. The development of such mappings has been the source of much controversy since the inception of the theory~\cite{steckelmacher1986knudsen}, leading to many lingering misconceptions (e.g., incorrect conductance values for square tubes). A further discussion of these misconceptions is outside of the scope of the presented research, instead, the simplified hydraulic diameter approximation is employed~\cite{ylilammi2018modeling}. That is, the diameter $d$ in \eqref{eq:diffusion} and \eqref{eq:knudsen} is replaced with $d \to h_d \cdot d$, where $h_d$ is the hydraulic diameter factor and $d$ is a relevant physical dimension. For example, for a wide rectangular trench with opening $d$ (c.f. Fig.~\ref{fig:geometry}), $h_d$ is estimated to be $2$~\cite{cremers2019conformality, ylilammi2018modeling}.

Equation~\eqref{eq:knudsen} is valid when reactant-wall collisions are more likely than reactant-reactant collisions (i.e., Knudsen number $\mathrm{Kn} > 10$). Should the rate of particle-particle collisions be comparable, i.e., $1 < \mathrm{Kn} < 10$, $D$ can be approximated with the Bosanquet interpolation formula~\cite{pollard1948gaseous}:
\begin{equation}
    \label{eq:bosanquet}
    \frac{1}{D} \approx \frac{1}{D_A} + \frac{1}{D_\mathrm{Kn}}
\end{equation}
In \eqref{eq:bosanquet}, $D_A$ is the conventional Chapman-Enskog gas-phase diffusivity~\cite{chapman1990mathematical} calculated from the particle hard-sphere diameter $d_A$ ($\mathrm{pm}$). In this work, we assume only Knudsen diffusivity ($D_A \to \infty$), except when otherwise indicated.

\begin{figure}[h!]
    \centering 
    \includegraphics[width=0.6\columnwidth]{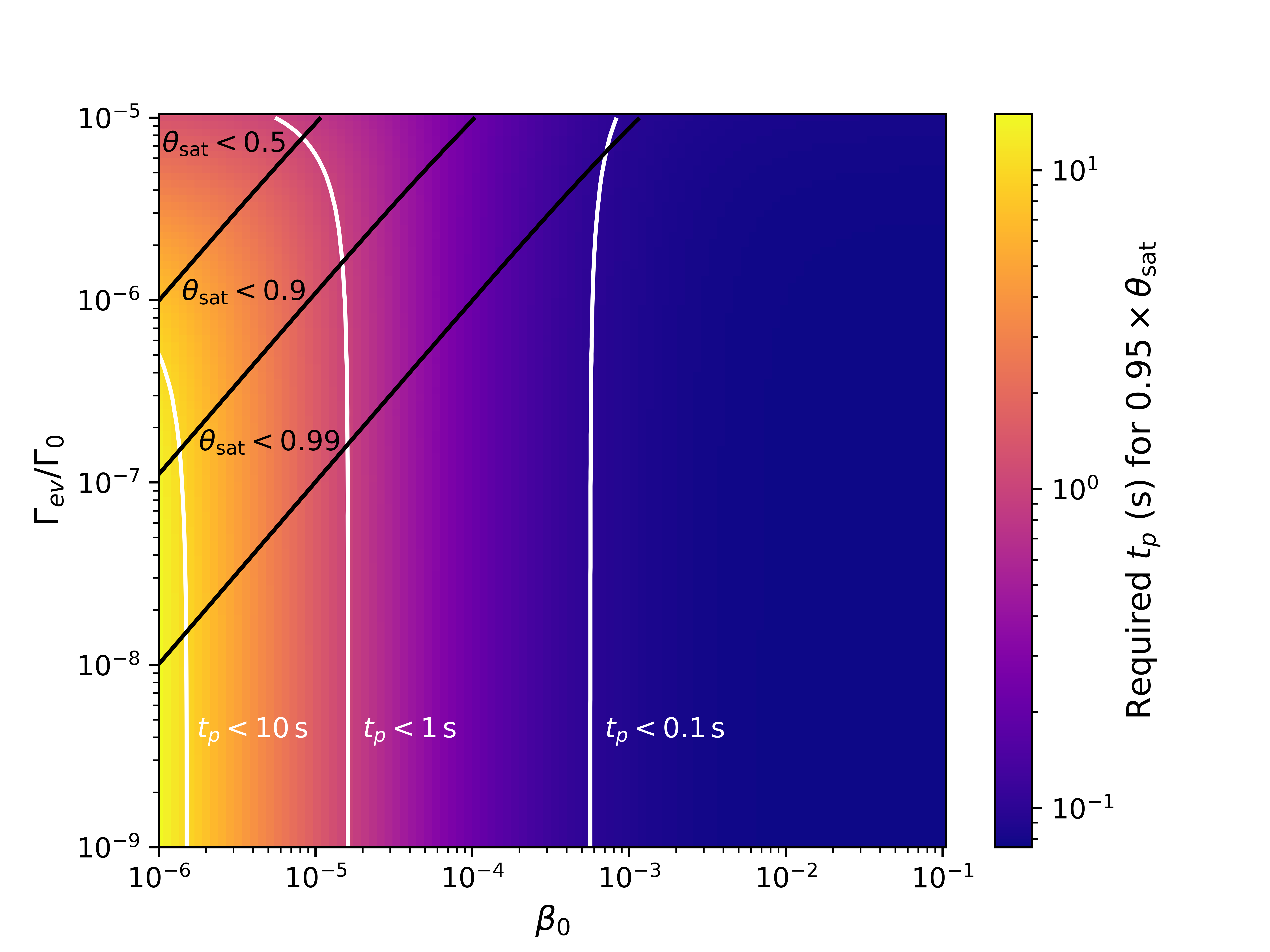}
    \caption{Impact of model parameters in the required $t_p$ to reach $95\%$ saturation at the bottom of a cylindrical structure with $d=1\,\mu\mathrm{m}$, $L=100\,\mu\mathrm{m}$ in a fictitious chemistry with $s_0 = 2\times10^{-19}\,\mathrm{m^2}$ and $\Gamma_0 = 10^{24}\,\mathrm{m^{-2}\,s^{-1}}$.}  
    \label{fig:heatmap}
\end{figure}

In Fig.~\ref{fig:heatmap}, the impact of the parameters $\beta_0$ and $\Gamma_\mathrm{ev}$ in the required $t_p$ for achieving $95\%$ saturation is presented. We define saturation as the final state of the surface coverage in the steady state, i.e., $d\theta_\mathrm{sat}(\vec{r})/dt = 0$. Since this situation is only reached on the limit $t_p \to \infty$, we identify a relevant time as the pulse time necessary to reach $0.95\cdot \theta_\mathrm{sat}$. In addition, as $\theta_\mathrm{sat}$ is defined on the entire structure, the coverage value at the bottom of the structure, i.e., $\theta_\mathrm{sat}(z{=}L)$, is chosen as the representative value for analysis, since it is where the impact of the flux restriction is the largest.

From Fig.~\ref{fig:heatmap}, we observe that $\beta_0$ has the most influence on the saturation time. Instead of directly impacting $t_p$, $\Gamma_\mathrm{ev}$ greatly affects the maximum coverage achievable at the  bottom of the structure. Therefore, it strictly limits the maximum aspect ratio achievable by a certain reactor configuration and must be considered in the design of novel technologies. In addition, previous work has shown that a high value of $\Gamma_\mathrm{ev}$ can impact the thickness profile particularly in the transition between a region with high growth to one with low growth~\cite{yim2022conformality}.

\subsection{Topography simulation}
\label{ssec:topo}

\begin{figure}[h!]
    \centering 
    \includegraphics[width=0.6\columnwidth]{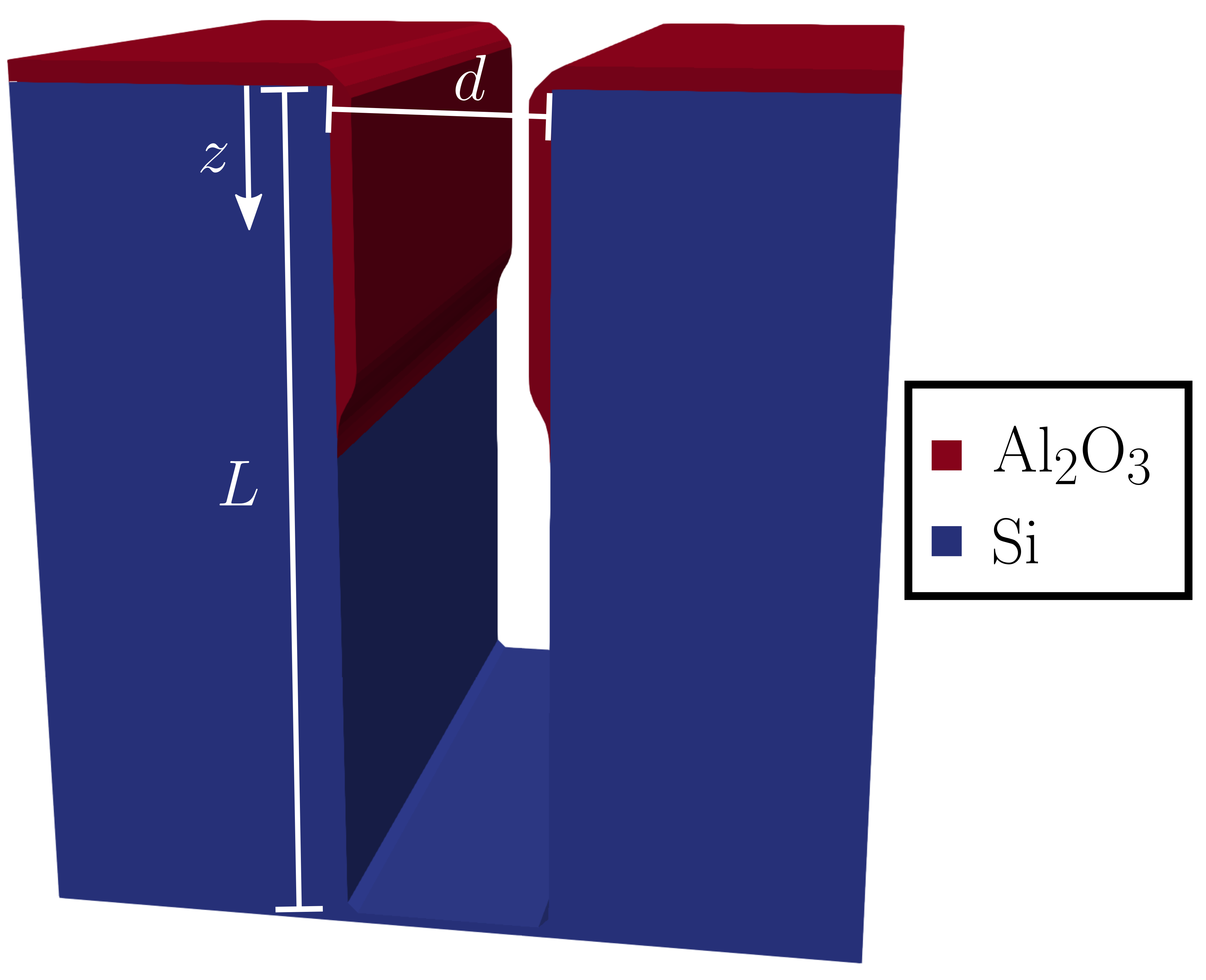}
    \caption{Illustration of simulated trench after ALD with incomplete conformality.}  
    \label{fig:geometry}
\end{figure}

In order to calculate the time evolution of the thickness profiles during the fabrication process in a manner compatible with simulation of further processing steps and DTCO, we employ the established level-set method~\cite{sethian1999level,klemenschits2018modeling} as implemented in \textit{ViennaLS}~\cite{ViennaLS} and in Silvaco's \textit{Victory Process}~\cite{VictoryProcess}. In this method, the surface is described as the zero level-set of a 3D function $\phi(\vec{r})$ which evolves in time according to the level-set equation
\begin{equation}
    \label{eq:levelset}
    \frac{\partial \phi (\vec{r}, t)}{\partial t} + V(\vec{r})\left|\nabla \phi(\vec{r}, t)\right| = 0,
\end{equation}
where $V(\vec{r})$ is a scalar velocity field describing how the surface should evolve over time, i.e., how a material is grown or etched. An illustration of a simulated 3D trench geometry after ALD of Al$_2$O$_3$ is shown in Fig.~\ref{fig:geometry}.

The methodology presented in Section~\ref{ssec:knudsenlangmuir} is limited to calculating $\theta(\vec{r})$. However, it is not straightforward to map $\theta(\vec{r})$ into $V(\vec{r})$. Growth rates can be calculated cycle-by-cycle by evolving the surface by the molecular layer thickness~\cite{filipovic2019modeling}, however, this imposes a performance penalty since the grid resolution must be small enough to capture the individual molecular layer and $\theta(\vec{r})$ must be calculated $N_\mathrm{cycles}$ times. This calculation repeats even though the geometry changes minimally between sequential cycles.

In order to capture a realistic ALD process with hundreds or thousands of cycles, a more efficient approach is required, bundling multiple cycles into the surface evolution step. For this, we introduce an artificial time $t^* = N_\mathrm{cycles}/C$ where $C$ is a numerical constant. It is important to note that $t^*$ is unrelated to $t_p$, since the latter is only required for the calculation of $\theta(\vec{r})$. In essence, to maintain unit consistency with \ref{eq:levelset}, the time variable $t^*$ represents a bundle of multiple ALD cycles. Thus, the velocity field becomes
\begin{equation}
    \label{eq:velocity}
    V(\vec{r}) = V(z) = C\cdot\mathrm{GPC}\cdot\theta(z)\text{.}
\end{equation}

The constant $C$ can be chosen by considering the involved number of cycles such that $t^* \approx 1$. In fact, the choice of $C$ plays a limited role in the determination of the bundling. Instead, in the involved level-set based topography simulators, $V(\vec{r})$ is assumed to be constant during each time step $\Delta t$ of the solution of \eqref{eq:levelset}. According to the Courant-Friedrichs-Lewy (CFL) condition~\cite{sethian1999level}, the time step is limited to allow an evolution of at most one grid spacing $\Delta x$, i.e., $\Delta t < \Delta x/\max{|V(\vec{r})|}$. After each time step, the geometrical inputs of \eqref{eq:diffusion}, namely $d$ and $L$, are updated. Thus, for \eqref{eq:velocity} to be physically meaningful, $\Delta x$ must be small enough such that the a change in the geometry of its magnitude does not significantly impact $\theta(\vec{r})$.

\section{Results}
\label{sec:results}

\subsection{The H$_2$O step: Temperature dependence}
\label{ssec:h2o}

\begin{figure}[h!]
    \centering 
    \includegraphics[width=0.6\columnwidth]{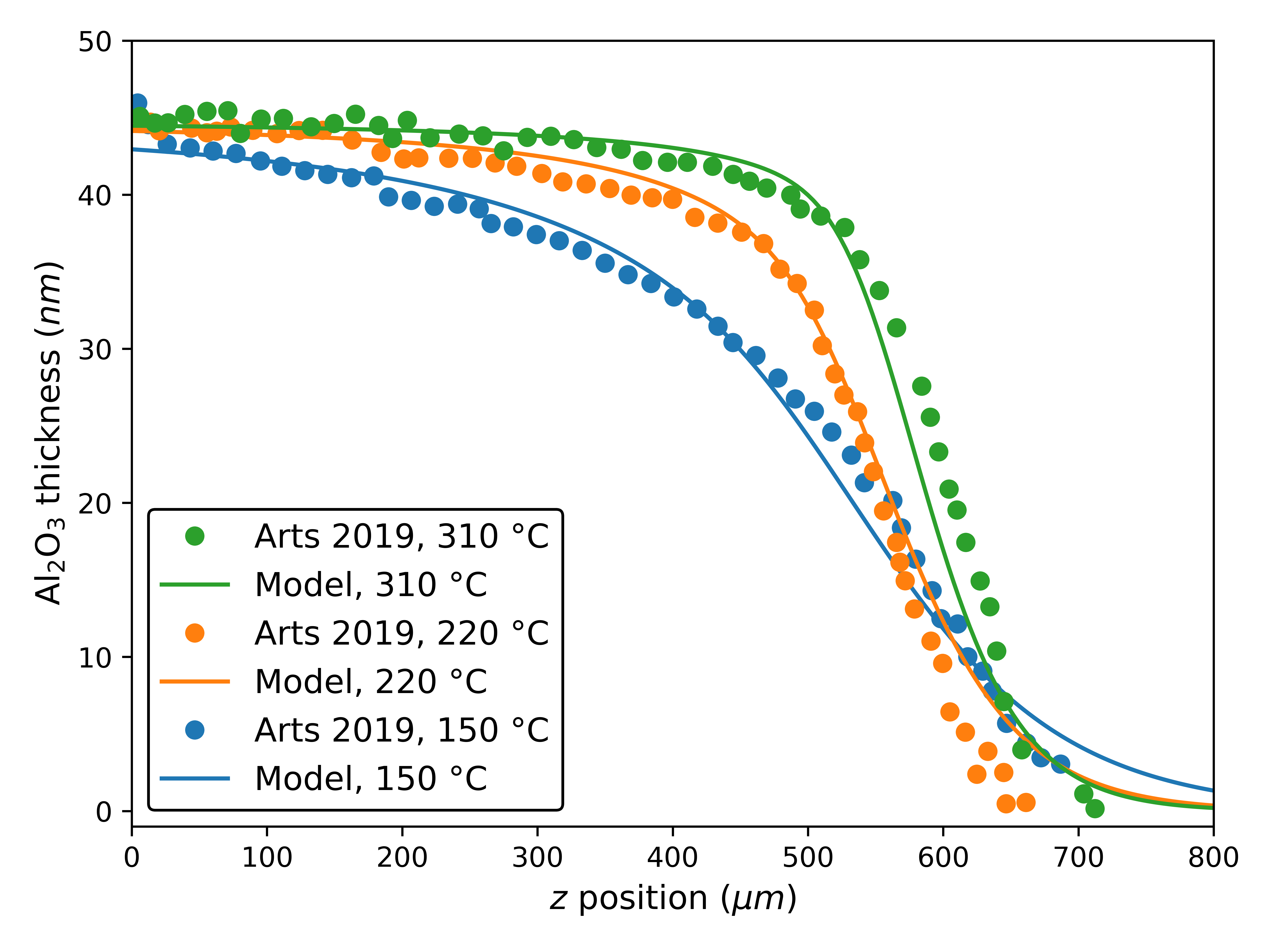}
    \caption{Comparison of topography simulation using the combined surface coverage model with the parameters from Table~\ref{tab:h2o} to H$_2$O-limited thickness profiles measured by Arts \textit{et al.}~\cite{arts2019sticking}}  
    \label{fig:h2o}
\end{figure}

We calibrate our model to measured thickness profiles of Al$_2$O$_3$ in the H$_2$O-limited regime. Arts \textit{et al.}~\cite{arts2019sticking} report film thicknesses in lateral HAR trench-like structures ($d=0.5\,\mu\mathrm{m}$, $L=5000\,\mu\mathrm{m}$) with an H$_2$O dose of approximately $750\,\mathrm{mTorr}{\cdot}\mathrm{s}$ after $400$ ALD cycles with a GPC of $1.12\,\text{\r{A}}$ at three calibrated substrate temperatures $T$ ($150\,\mathrm{^\circ C}$, $220\,\mathrm{^\circ C}$, and $310\,\mathrm{^\circ C}$). We estimate $t_p$ to be $0.1\,\mathrm{s}$. We were unable to reproduce the reported penetration depths with realistic values of density~\cite{ylivaara2014aluminum} in the calculation of $s_0$~\cite{ylilammi2018modeling} using the ``billiard ball" approximation. Therefore, we treat $s_0$ as another parameter to be estimated, for which we obtain the value of $3.36\cdot 10^{-19}\,\mathrm{m^2}$. The parameters for each $T$ were obtained by manual adjustment and visual comparison to the experimental data and are provided in Table~\ref{tab:h2o}. The model comparison to experimental data is given in Fig.~\ref{fig:h2o}. The authors of the original work also estimate $\beta_0$ from the slope at $50\%$ height, and those values are reported in Table~\ref{tab:h2o}.

\begin{table}[h]
        \centering
        \caption{Model parameters for the H$_2$O step of ALD of Al$_2$O$_3$ calibrated to measurements from~\cite{arts2019sticking}.}
        \label{tab:h2o}
        \begin{tabular}{llll}
        \toprule
        \textbf{Parameter} & $150\,\mathrm{^\circ C}$ & $220\,\mathrm{^\circ C}$ & $310\,\mathrm{^\circ C}$ \\ \midrule
        $\Gamma_\mathrm{ev}$ ($\mathrm{m^{-2}s^{-1}}$) & $6.5 \cdot 10^{19}$ & $5.0 \cdot 10^{19}$ & $3.5 \cdot 10^{19}$\\
        \hline
        $\beta_0$ & $5.0 \cdot 10^{-5}$ & $1.2 \cdot 10^{-4}$ & $1.9 \cdot 10^{-4}$\\
        \hline
        \begin{tabular}{@{}c@{}} {} \\[0ex]  $\beta_0$, estimated range from~\cite{arts2019sticking} \\[0ex] {} \end{tabular} & 
        \begin{tabular}{@{}c@{}} $1.4 \cdot 10^{-5}$ \\[0ex] $-$ \\[0ex] $2.3 \cdot 10^{-5}$\end{tabular} & \begin{tabular}{@{}c@{}} $0.8 \cdot 10^{-4}$ \\[0ex] $-$ \\[0ex] $2.0 \cdot 10^{-4}$\end{tabular} & \begin{tabular}{@{}c@{}} $0.9 \cdot 10^{-4}$ \\[0ex] $-$ \\[0ex] $2.5 \cdot 10^{-4}$\end{tabular}\\\bottomrule
        \end{tabular}
\end{table}

In Fig.~\ref{fig:h2o}, we note a good agreement between our topography simulations using the combined model for surface coverage and the reported experimental profiles. The estimated values of $\beta_0$ are also generally consistent with the estimated ranges from the original work, which is expected since it also relies on first-order Langmuir kinetics. However, we expect that our methodology provides a more accurate estimate, including on the discrepant value at $150\,\mathrm{^\circ C}$, since we consider the entire profile and we include $\Gamma_\mathrm{ev}$. Nonetheless, it is possible that we overestimate $\Gamma_\mathrm{ev}$ since we do not consider the purge step. The reduction in thickness and the less abrupt transition between the region with high growth for that profile is strong evidence of the important role of reversible reactions, which is supported by other modeling studies~\cite{yim2022conformality}.

\begin{figure}[h!]
    \centering 
    \includegraphics[width=0.6\columnwidth]{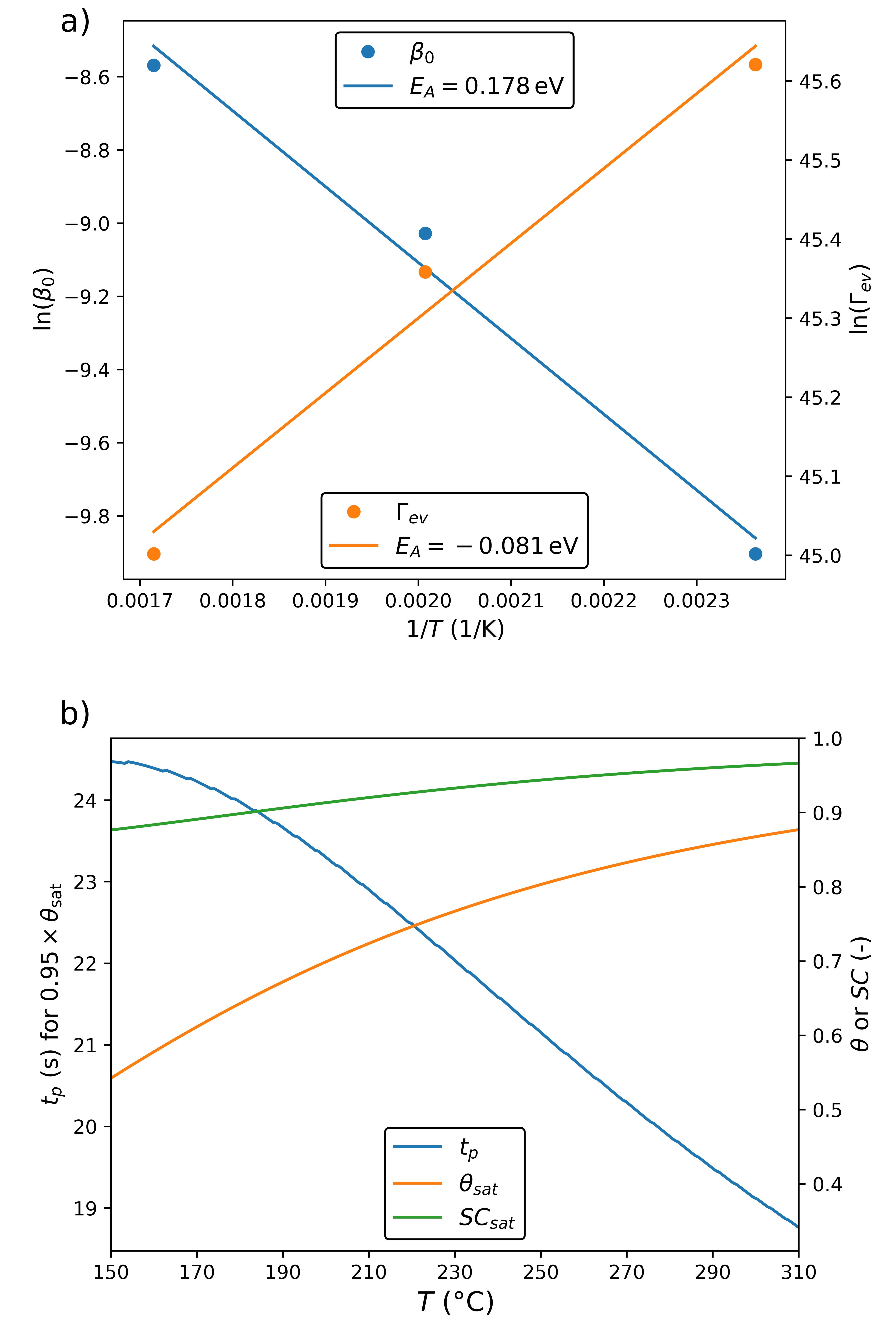}
    \caption{(a) Arrhenius analysis of $\beta_0$ and $\Gamma_\mathrm{ev}$ from Table~\ref{tab:h2o}. (b) After parameterization to $T$, its effect is investigated in the required $t_p$ to reach $95\%$ of saturation, $\theta_\mathrm{sat}$ at $z=L$ and $\mathrm{SC}_\mathrm{sat}$.}  
    \label{fig:arrhenius}
\end{figure}

Due to the availability of data at different substrate temperatures, we perform an indicative Arrhenius analysis, shown in Fig.~\ref{fig:arrhenius}. In Fig.~\ref{fig:arrhenius} (a), we observe that the $\beta_0$ increases and $\Gamma_\mathrm{ev}$ decreases with increasing $T$. This suggests that the increase in temperature not only makes the reaction more efficient but also, counter-intuitively, reduces the reversibility of the reaction. This is the cause of the negative value of $E_A$ on the Arrhenius fit of $\Gamma_\mathrm{ev}$, which is not itself the true activation energy of the reaction. Instead, we interpret the value of $E_A$ from the linear fit of $\beta_0$ ($0.178\,\mathrm{eV}$) as that representing the energy barrier involved in the reaction, since it is the one which must be overcome on a clean surface (i.e., $\theta{=}0$). Although this value is lower than first-principle studies suggest ($1.101 \, \mathrm{eV}$)~\cite{seo2018molecular}, it is consistent with a recent experimental analysis exploring a two-stage reaction, where $E_A$ is estimated as $0.166 \pm 0.02\,\mathrm{eV}$~\cite{sperling2020atomic}. This two-stage reaction is a possible cause of failure of the ``billiard ball" approximation for $s_0$.

From the fitted Arrhenius relationships, both model parameters ($\beta_0$ and $\Gamma_\mathrm{ev}$) can be expressed as functions of the single physical variable $T$. Thus, the parameter analysis from Fig.~\ref{fig:heatmap} can be reduced from three to two dimensions, as shown in Fig.~\ref{fig:arrhenius} (b). We observe that the saturation $t_p$ reduces and $\theta_\mathrm{sat}$ at $z=L$ increases with higher temperatures, as is expected from a more thermodynamically favorable reaction. However, in many experimental situations $\theta$ is not easily measurable. Instead, the step coverage ($\mathrm{SC}$) is commonly measured~\cite{yanguas2016growth}. After saturation, i.e. $d\theta/dt=0$, we estimate the step coverage to be $\mathrm{SC}_\mathrm{sat}=\theta_\mathrm{sat}({z=L})/\theta_\mathrm{sat}({z=0})$. Interestingly, we note that $\mathrm{SC}_\mathrm{sat}$ is high and nearly constant for the entire tested temperature range even though $\theta_\mathrm{sat}$ has a larger variation.
Thus, we expect that, at low temperatures, the film quality could be low due to the presence of defects such as vacancies and voids.

\subsection{The TMA step and geometric parameters}
\label{ssec:loading}

Similarly to Section~\ref{ssec:h2o}, the model is manually calibrated to published thickness profiles of Al$_2$O$_3$ in the TMA-limited regime. Due to the comparatively higher complexity of TMA, this step has received more research attention, therefore, we are able to simultaneously apply our model to multiple independent experiments in similar lateral HAR structures ($d=0.5\,\mu\mathrm{m}$)~\cite{ylilammi2018modeling,arts2019sticking,yim2020saturation}. All available reactor and film parameters were taken directly from the original publications. The unavailable data was estimated as follows: For Ylilammi \textit{et al.}~\cite{ylilammi2018modeling} (and footnotes from~\cite{yim2020saturation}), we estimate $p_A = 325\,\mathrm{mTorr}$; for Arts \textit{et al.}~\cite{arts2019sticking}, $t_p = 0.4\,\mathrm{s}$ and $\rho_\mathrm{Al_2O_3}=3000\,\mathrm{kg}/\mathrm{m^3}$; and for Yim and Ylivaara \textit{et al.}~\cite{yim2020saturation}, $p_A = 160\,\mathrm{mTorr}$.

Since all reported thickness profiles were obtained on a restricted range of set temperatures ($275\,^\circ\mathrm{C}$ in~\cite{arts2019sticking}, $300\,^\circ\mathrm{C}$ otherwise), we manually calibrate our model to all profiles with the same parameter set presented in Table~\ref{tab:tma}, including the estimates of $\beta_0$ from the original works. The disparity is likely due to the effect of $\Gamma_\mathrm{ev}$, which is corroborated by the most similar value being that from~\cite{ylilammi2018modeling}, whose approach also considers reversible kinetics.

\begin{figure}[h!]
    \centering 
    \includegraphics[width=0.6\columnwidth]{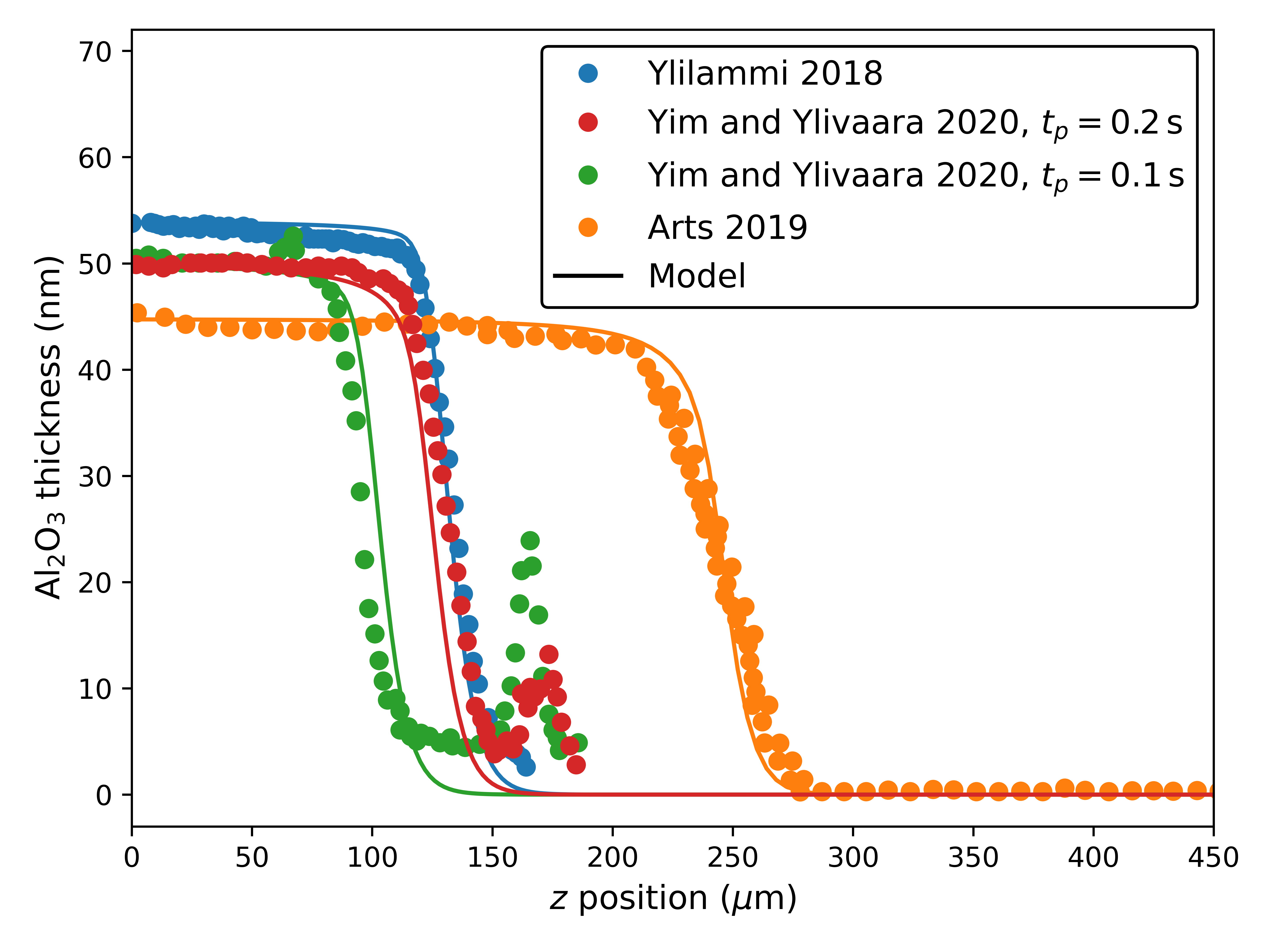}
    \caption{Comparison of simulation with parameters from Table~\ref{tab:tma} to TMA-limited thickness profiles reported by Ylilammi \textit{et al.}~\cite{ylilammi2018modeling}, Arts \textit{et al.}~\cite{arts2019sticking}, and Yim and Ylivaara \textit{et al.}~\cite{yim2020saturation}.}  
    \label{fig:tma}
\end{figure}

The comparison to the published measured profiles is provided in Fig.~\ref{fig:tma}, showing good agreement. This is strong evidence for the hypothesis discussed in Section~\ref{ssec:knudsenlangmuir} that the model parameters are determined by the reactor setup, most importantly the reactor $T$. The peaks shown in the experimental data from~\cite{yim2020saturation} are disregarded since they are reported to be spurious interactions with the pillars sustaining the structure.

\begin{table}[h]
        \centering
        \caption{Model parameters for the TMA step of ALD of Al$_2$O$_3$ calibrated to multiple measurements~\cite{ylilammi2018modeling,arts2019sticking,yim2020saturation}.}
        \label{tab:tma}
        \begin{tabular}{lllll}
        \toprule
        $\Gamma_\mathrm{ev}$ ($\mathrm{m^{-2}s^{-1}}$) & $\beta_0$ & $\beta_0$ from~\cite{ylilammi2018modeling} & $\beta_0$ from~\cite{arts2019sticking} & $\beta_0$ from~\cite{yim2020saturation}\\ \midrule
         $3.0 \cdot 10^{19}$ & $7.5 \cdot 10^{-3}$ & $5.7 \cdot 10^{-3}$ & $(0.5{-}2.0) \cdot 10^{-3}$ & $4.0 \cdot 10^{-3}$ \\ \bottomrule
        \end{tabular}
        \end{table}
        
\begin{figure}[h!]
    \centering 
    \includegraphics[width=0.6\columnwidth]{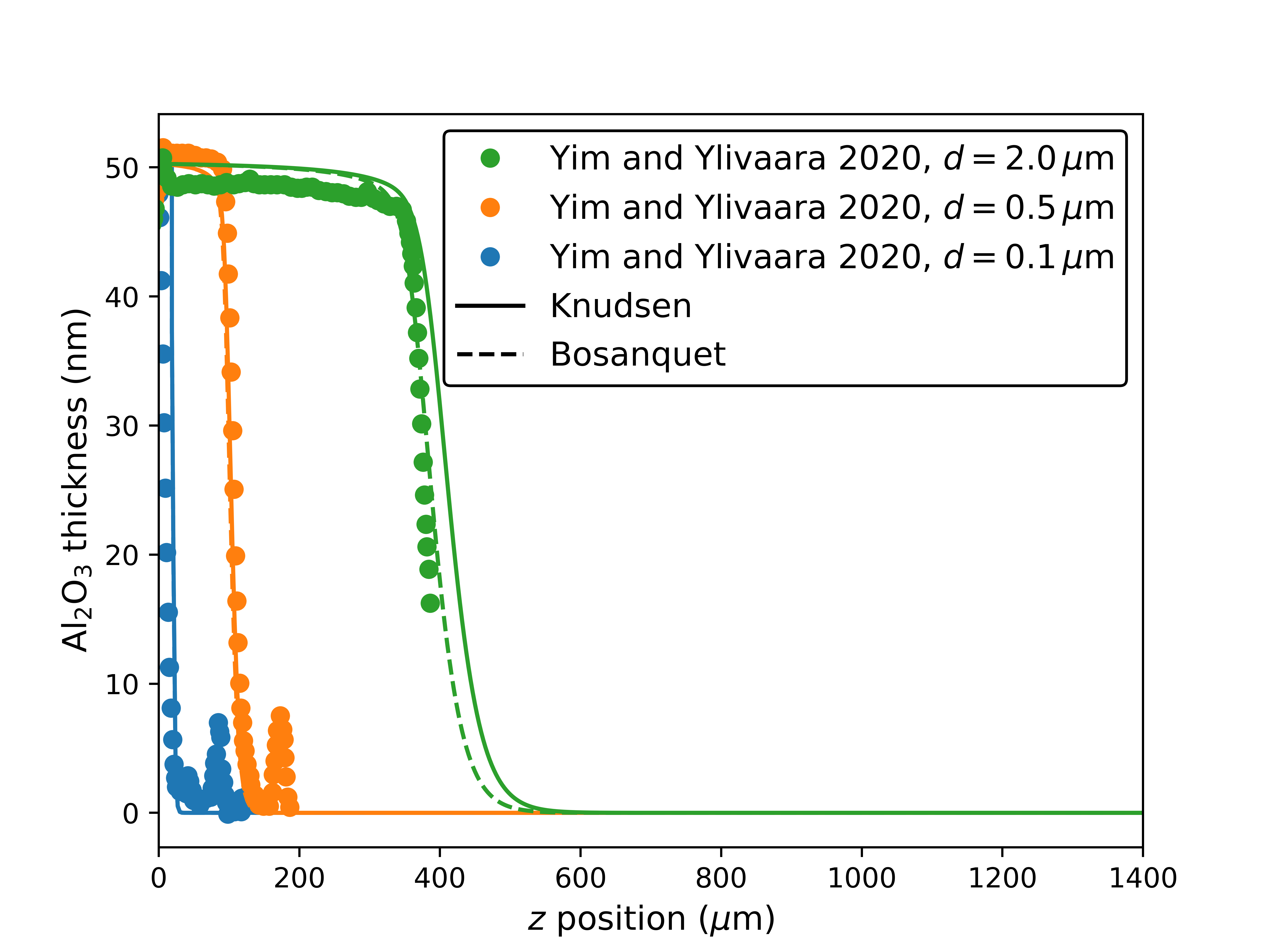}
    \caption{Comparison of simulated structures to profiles reported by Yim and Ylivaara~\textit{et al.}~\cite{yim2020saturation} for lateral HAR structures with different initial openings $d$ using parameters from Table~\ref{tab:tma}. ``Knudsen" shows the model using only Knudsen diffusivity, while ``Bosanquet" includes gas-phase diffusivity.}  
    \label{fig:vary_geo}
\end{figure}

We reproduce additional experiments by Yim and Ylivaara \textit{et al.}~\cite{yim2020saturation} in lateral HAR structures with different initial openings $d$, shown in Fig.~\ref{fig:vary_geo}. The discrepancy in the structure with $d=0.1\,\mu\mathrm{m}$ is due to the limits of our model when the opening becomes fully constricted. In this situation, the approximation that the entire geometry can be represented by an evolving but single value of $d$ starts to fail. One additional limitation is the failure of the hydraulic diameter approximation in a constricted structure, since it is not rigorously justified and has significant discrepancies with regards to established results~\cite{steckelmacher1986knudsen}. For the structure with opening $d=2.0\,\mu\mathrm{m}$, pure Knudsen diffusivity is no longer valid, since $\mathrm{Kn} \approx 8.9$. We recover accuracy by using \eqref{eq:bosanquet} (marked ``Bosanquet") which is calculated using the hard-sphere diameters of TMA $d_\mathrm{TMA} = 591\,\mathrm{pm}$ and of nitrogen (N$_2$, the carrier gas) $d_\mathrm{N_2} = 374\,\mathrm{pm}$~\cite{ylilammi2018modeling}.

\section{Summary and Outlook}
\label{ssec:concl}

In this work, we present a surface coverage model for incomplete conformality during ALD in HAR structures based on diffusive particle transport and reversible first-order Langmuir kinetics which combines insights from multiple established modeling approaches. By focusing on the evaporation flux, we achieve a good fit to experimental data and also obtain further chemical insights from the saturation behavior. Also, by approximating the diffusivity with the Bosanquet formula, we are able to capture processing conditions with lower Knudsen numbers. Finally, we present an approach for efficiently integrating our model with a level-set topography simulator by bundling multiple ALD cycles into an artificial time unit.

We manually calibrate our model to reported thickness profiles in the prototypical ALD of Al$_2$O$_3$ from H$_2$O and TMA. We study the impact of temperature in H$_2$O-limited profiles, indicating the strong impact of the evaporation flux at lower temperatures and extracting an activation energy of $0.178\,\mathrm{eV}$ which is comparable with recent experimental studies. From calibrating our simulation with a single parameter set to multiple independent experiments in the TMA-limited regime, we strengthen the hypothesis that the parameters are strongly related to the reactor condition, most importantly to its temperature.

Our ALD modeling can be further improved by integrating a more accurate flux calculation methodology such as Monte Carlo ray tracing and additional physical phenomena, such as losses due to recombination, partial decomposition, and the effect of impurities. Since the evaporation flux plays such an important role, the explicit consideration of the purge step can further enhance the model. A more rigorous estimation approach would also enable estimation of the error bounds, which would improve the connections to experimental data. Finally, our robust level-set simulation approach enables the simulation of further processing steps and of process-aware device operation and could be applied to, e.g., atomic layer etching for 3D integration of novel memories.

\section*{Acknowledgments}
The financial support by the Austrian Federal Ministry for Digital and Economic Affairs, the National Foundation for Research, Technology and Development and the Christian Doppler Research Association is gratefully acknowledged. This work was supported in part by the Austrian Research Promotion Agency FFG under Project 878662 PASTE-DTCO.

\bibliographystyle{elsarticle-num} 
\bibliography{cas-refs}

\begin{thebibliography}{10}
\expandafter\ifx\csname url\endcsname\relax
  \def\url#1{\texttt{#1}}\fi
\expandafter\ifx\csname urlprefix\endcsname\relax\def\urlprefix{URL }\fi
\expandafter\ifx\csname href\endcsname\relax
  \def\href#1#2{#2} \def\path#1{#1}\fi

\bibitem{cremers2019conformality}
V.~Cremers, R.~L. Puurunen, J.~Dendooven, Conformality in atomic layer
  deposition: {C}urrent status overview of analysis and modelling, Appl Phys
  Rev 6~(2) (2019) 021302.
\newblock \href {https://doi.org/10.1063/1.5060967}
  {\path{doi:10.1063/1.5060967}}.

\bibitem{knoops2015atomicbook}
H.~Knoops, S.~Potts, A.~Bol, W.~Kessels, 27 - {A}tomic {L}ayer {D}eposition,
  in: T.~F. Kuech (Ed.), Handbook of Crystal Growth, {S}econd Edition,
  North-Holland, 2015, pp. 1101--1134.
\newblock \href {https://doi.org/10.1016/B978-0-444-63304-0.00027-5}
  {\path{doi:10.1016/B978-0-444-63304-0.00027-5}}.

\bibitem{jakschik2004physical}
S.~Jakschik, U.~Schroeder, T.~Hecht, G.~Dollinger, A.~Bergmaier, J.~Bartha,
  Physical properties of {ALD}-{Al}$_2${O}$_3$ in a {DRAM}-capacitor equivalent
  structure comparing interfaces and oxygen precursors, Mater Sci Eng B 107~(3)
  (2004) 251--254.
\newblock \href {https://doi.org/10.1016/j.mseb.2003.09.044}
  {\path{doi:10.1016/j.mseb.2003.09.044}}.

\bibitem{fischer2022control}
A.~Fischer, A.~Routzahn, R.~J. Gasvoda, J.~Sims, T.~Lill, Control of etch
  profiles in high aspect ratio holes via precise reactant dosing in thermal
  atomic layer etching, J Vac Sci Technol A 40~(2) (2022) 022603.
\newblock \href {https://doi.org/10.1116/6.0001691}
  {\path{doi:10.1116/6.0001691}}.

\bibitem{puurunen2005surface}
R.~L. Puurunen, Surface chemistry of atomic layer deposition: {A} case study
  for the trimethylaluminum/water process, J Appl Phys 97~(12) (2005) 9.
\newblock \href {https://doi.org/10.1063/1.1940727}
  {\path{doi:10.1063/1.1940727}}.

\bibitem{gobbert2002predictive}
M.~K. Gobbert, V.~Prasad, T.~S. Cale, Predictive modeling of atomic layer
  deposition on the feature scale, Thin Solid Films 410~(1-2) (2002) 129--141.
\newblock \href {https://doi.org/10.1016/S0040-6090(02)00236-5}
  {\path{doi:10.1016/S0040-6090(02)00236-5}}.

\bibitem{gordon2003kinetic}
R.~G. Gordon, D.~Hausmann, E.~Kim, J.~Shepard, A kinetic model for step
  coverage by atomic layer deposition in narrow holes or trenches, Chem Vap
  Depos 9~(2) (2003) 73--78.
\newblock \href {https://doi.org/10.1002/cvde.200390005}
  {\path{doi:10.1002/cvde.200390005}}.

\bibitem{yanguas2012self}
A.~Yanguas-Gil, J.~W. Elam, Self-limited reaction-diffusion in nanostructured
  substrates: Surface coverage dynamics and analytic approximations to ald
  saturation times, Chem Vap Depos 18~(1-3) (2012) 46--52.
\newblock \href {https://doi.org/10.1002/cvde.201106938}
  {\path{doi:10.1002/cvde.201106938}}.

\bibitem{ylilammi2018modeling}
M.~Ylilammi, O.~M. Ylivaara, R.~L. Puurunen, Modeling growth kinetics of thin
  films made by atomic layer deposition in lateral high-aspect-ratio
  structures, J Appl Phys 123~(20) (2018) 205301.
\newblock \href {https://doi.org/10.1063/1.5028178}
  {\path{doi:10.1063/1.5028178}}.

\bibitem{szmyt2021atomic}
W.~Szmyt, C.~Guerra-Nu{\~n}ez, L.~Huber, C.~Dransfeld, I.~Utke, Atomic layer
  deposition on porous substrates: {F}rom general formulation to fibrous
  substrates and scaling laws, Chem Mater 34~(1) (2021) 203--216.
\newblock \href {https://doi.org/10.1021/acs.chemmater.1c03164}
  {\path{doi:10.1021/acs.chemmater.1c03164}}.

\bibitem{schwille2017temperature}
M.~C. Schwille, T.~Sch{\"o}ssler, F.~Sch{\"o}n, M.~Oettel, J.~W. Bartha,
  Temperature dependence of the sticking coefficients of bis-diethyl
  aminosilane and trimethylaluminum in atomic layer deposition, J Vac Sci
  Technol A 35~(1) (2017) 01B119.
\newblock \href {https://doi.org/10.1116/1.4971197}
  {\path{doi:10.1116/1.4971197}}.

\bibitem{poodt2017effect}
P.~Poodt, A.~Mameli, J.~Schulpen, W.~Kessels, F.~Roozeboom, Effect of reactor
  pressure on the conformal coating inside porous substrates by atomic layer
  deposition, J Vac Sci Technol A 35~(2) (2017) 021502.
\newblock \href {https://doi.org/10.1116/1.4973350}
  {\path{doi:10.1116/1.4973350}}.

\bibitem{arts2019sticking}
K.~Arts, V.~Vandalon, R.~L. Puurunen, M.~Utriainen, F.~Gao, W.~M. Kessels,
  H.~C. Knoops, Sticking probabilities of h$_2$o and {A}l({CH}$_3$)$_3$ during
  atomic layer deposition of {A}l$_2${O}$_3$ extracted from their impact on
  film conformality, J Vac Sci Technol A 37~(3) (2019) 030908.
\newblock \href {https://doi.org/10.1116/1.5093620}
  {\path{doi:10.1116/1.5093620}}.

\bibitem{TUW-297930}
X.~Klemenschits, S.~Selberherr, L.~Filipovic, {C}ombined process simulation and
  simulation of an {S}{R}{A}{M} cell of the 5nm technology node, in:
  {P}roceedings of the {I}nternational {C}onference on {S}imulation of
  {S}emiconductor {P}rocesses and {D}evices ({S}{I}{S}{P}{A}{D}), 2021, pp.
  23--27.
\newblock \href {https://doi.org/10.1109/SISPAD54002.2021.9592605}
  {\path{doi:10.1109/SISPAD54002.2021.9592605}}.

\bibitem{filipovic2019modeling}
L.~Filipovic, Modeling and simulation of atomic layer deposition, in:
  {P}roceedings of the {I}nternational {C}onference on {S}imulation of
  {S}emiconductor {P}rocesses and {D}evices ({S}{I}{S}{P}{A}{D}), IEEE, 2019,
  pp. 323--326.
\newblock \href {https://doi.org/10.1109/SISPAD.2019.8870462}
  {\path{doi:10.1109/SISPAD.2019.8870462}}.

\bibitem{yim2022conformality}
J.~Yim, E.~Verkama, J.~A. Velasco, K.~Arts, R.~L. Puurunen, Conformality of
  atomic layer deposition in microchannels: {I}mpact of process parameters on
  the simulated thickness profile, Phys Chem Chem Phys 24~(15) (2022)
  8645--8660.
\newblock \href {https://doi.org/10.1039/d1cp04758b}
  {\path{doi:10.1039/d1cp04758b}}.

\bibitem{pollard1948gaseous}
W.~Pollard, R.~D. Present, On gaseous self-diffusion in long capillary tubes,
  Phys Rev 73~(7) (1948) 762.
\newblock \href {https://doi.org/10.1103/PhysRev.73.762}
  {\path{doi:10.1103/PhysRev.73.762}}.

\bibitem{sethian1999level}
J.~A. Sethian, Level set methods and fast marching methods, {S}econd Edition,
  Cambridge University Press, 1999.

\bibitem{klemenschits2018modeling}
X.~Klemenschits, S.~Selberherr, L.~Filipovic, Modeling of gate stack patterning
  for advanced technology nodes: {A} review, Micromach 9~(12) (2018) 631.
\newblock \href {https://doi.org/10.3390/mi9120631}
  {\path{doi:10.3390/mi9120631}}.

\bibitem{ViennaLS}
{ViennaLS}, {A}vailable online: \url{https://viennatools.github.io/ViennaLS}
  (accessed 08 June 2022).

\bibitem{VictoryProcess}
Silvaco, {V}ictory {P}rocess, {A}vailable online:
  \url{www.silvaco.com/tcad/victory-process-3d/} (accessed 08 June 2022).

\bibitem{keuter2015modeling}
T.~Keuter, N.~H. Menzler, G.~Mauer, F.~Vondahlen, R.~Va{\ss}en, H.~P.
  Buchkremer, Modeling precursor diffusion and reaction of atomic layer
  deposition in porous structures, J Vac Sci Technol A 33~(1) (2015) 01A104.
\newblock \href {https://doi.org/10.1116/1.4892385}
  {\path{doi:10.1116/1.4892385}}.

\bibitem{yanguas2021reactor}
A.~Yanguas-Gil, J.~A. Libera, J.~W. Elam, Reactor scale simulations of {ALD}
  and {ALE}: {I}deal and non-ideal self-limited processes in a cylindrical and
  a 300 mm wafer cross-flow reactor, J Vac Sci Technol A 39~(6) (2021) 062404.
\newblock \href {https://doi.org/10.1116/6.0001212}
  {\path{doi:10.1116/6.0001212}}.

\bibitem{fang2019atomic}
W.-Z. Fang, Y.-Q. Tang, C.~Ban, Q.~Kang, R.~Qiao, W.-Q. Tao, Atomic layer
  deposition in porous electrodes: {A} pore-scale modeling study, Chem Eng J
  378 (2019) 122099.
\newblock \href {https://doi.org/10.1016/j.cej.2019.122099}
  {\path{doi:10.1016/j.cej.2019.122099}}.

\bibitem{yanguas2016growth}
A.~Yanguas-Gil, {Growth and Transport in Nanostructured Materials: Reactive
  Transport in PVD, CVD, and ALD}, Springer, 2016.

\bibitem{chapman1990mathematical}
S.~Chapman, T.~G. Cowling, The mathematical theory of non-uniform gases,
  {T}hird Edition, Cambridge University Press, 1991.

\bibitem{Knudsen1909}
M.~Knudsen, {Eine Revision der Gleichgewichtsbedingung der Gase. Thermische
  Molekularströmung}, Annalen der Physik 336 (1909) 205--229.
\newblock \href {https://doi.org/10.1002/ANDP.19093360110}
  {\path{doi:10.1002/ANDP.19093360110}}.

\bibitem{steckelmacher1986knudsen}
W.~Steckelmacher, Knudsen flow 75 years on: {T}he current state of the art for
  flow of rarefied gases in tubes and systems, Rep Prog Phys 49~(10) (1986)
  1083.
\newblock \href {https://doi.org/10.1088/0034-4885/49/10/001}
  {\path{doi:10.1088/0034-4885/49/10/001}}.

\bibitem{ylivaara2014aluminum}
O.~M. Ylivaara, X.~Liu, L.~Kilpi, J.~Lyytinen, D.~Schneider, M.~Laitinen,
  J.~Julin, S.~Ali, S.~Sintonen, M.~Berdova, et~al., Aluminum oxide from
  trimethylaluminum and water by atomic layer deposition: {T}he temperature
  dependence of residual stress, elastic modulus, hardness and adhesion, Thin
  Solid Films 552 (2014) 124--135.
\newblock \href {https://doi.org/10.1016/j.tsf.2013.11.112}
  {\path{doi:10.1016/j.tsf.2013.11.112}}.

\bibitem{seo2018molecular}
S.~Seo, T.~Nam, H.~Kim, B.~Shong, et~al., Molecular oxidation of
  surface--{CH}$_3$ during atomic layer deposition of {A}l$_2${O}$_3$ with
  {H}$_2${O}, {H}$_2${O}$_2$, and {O}$_3$: {A} theoretical study, Appl Surf Sci
  457 (2018) 376--380.
\newblock \href {https://doi.org/10.1016/j.apsusc.2018.06.160}
  {\path{doi:10.1016/j.apsusc.2018.06.160}}.

\bibitem{sperling2020atomic}
B.~A. Sperling, B.~Kalanyan, J.~E. Maslar, Atomic layer deposition of
  {A}l$_2${O}$_3$ using trimethylaluminum and {H}$_2${O}: {T}he kinetics of the
  {H}$_2${O} half-cycle, J Phys Chem C 124~(5) (2020) 3410--3420.
\newblock \href {https://doi.org/10.1021/acs.jpcc.9b11291}
  {\path{doi:10.1021/acs.jpcc.9b11291}}.

\bibitem{yim2020saturation}
J.~Yim, O.~M. Ylivaara, M.~Ylilammi, V.~Korpelainen, E.~Haimi, E.~Verkama,
  et~al., Saturation profile based conformality analysis for atomic layer
  deposition: {A}luminum oxide in lateral high-aspect-ratio channels, Phys Chem
  Chem Phys 22~(40) (2020) 23107--23120.
\newblock \href {https://doi.org/10.1039/d0cp03358h}
  {\path{doi:10.1039/d0cp03358h}}.

\end{thebibliography}





\end{document}